\documentclass[review ,12pt]{article}
\usepackage{afterpage}
\usepackage{float}
\usepackage{longtable}
\usepackage{psfrag}
\usepackage{lscape}
\usepackage{natbib}
\usepackage{setspace}
\usepackage{enumerate}
\usepackage{makeidx}
\usepackage{amsmath}    
\usepackage{graphicx}   
\usepackage{verbatim}   
\usepackage{color}      
\usepackage{subfigure}  
\usepackage{hyperref}   
\usepackage{chemarr}
\usepackage[svgnames]{xcolor}
\usepackage{chngcntr}

\newcommand{\pd}{\partial}

\def\longrightharpoonup{\relbar\joinrel\rightharpoonup}
\def\longleftharpoondown{\leftharpoondown\joinrel\relbar}

\def\longrightleftharpoons{
  \mathop{
    \vcenter{
      \hbox{
        \ooalign{
          \raise1pt\hbox{$\longrightharpoonup\joinrel$}\crcr
          \lower1pt\hbox{$\longleftharpoondown\joinrel$}
        }
      }
    }
  }
}

\title{A Vertical Architecture for Increasing Photogalvanic Solar Cell Efficiency: Theory and Modeling}

\author{Mohammad Ali Mahmoudzadeh * , John D.W. Madden \\
\small Department of Electrical, University of British Columbia,\\
\small 2332 Main Mall, Vancouver BC, V6T 1Z4, Canada \\
\small * Corresponding author: ali-m@ece.ubc.ca}

\begin{document}
\maketitle

\begin{abstract}
Photogalvanic solar cells, the original dye based solar cell, have yet to fulfill their promise as a low fabrication cost, scalable energy conversion system. The efficient performance of photogalvanic cells relies on high dye solubility and selective electrodes with fast electron transfer kinetics. A new configuration  is proposed for photogalvanic cells that removes these impractical requirements. Instead of illuminating the device through the electrode, as is the conventional approach, a new vertical configuration is employed with light coming between the two electrodes. This way, the light absorption and hence electron generation is spread through the depth of the device. The depth therefore can be adjusted according to the concentration of the dyes to absorb all the incoming photons even with low solubility dyes. As a result of distributed electron generation, unreasonably fast electrode kinetics are no longer  required. The proposed configuration is mathematically modeled and the advantages over the conventional cell are shown. A numerical model is built for more detailed analysis that gives practical guidelines for working towards device parameters with high power conversion efficiency. The readily available Thionine-Iron dye-mediator couple could achieve 6\% efficiency if highly selective electrodes are used, compared to 0.45\% at best using the conventional approach. The analysis suggests that upon the realization of highly selective electrodes and an improved dye/mediator couple, an efficiency of 13\%, and potentially higher, should be achievable from the new configuration.
\end{abstract}

%

\section{Introduction}
Solar energy is the most abundant of the readily available renewable energy sources. So far, the cost of conventional solar power relative to fossil fuel alternatives has impeded its widespread use in grid-tied locations. Many approaches are being taken to reduce the cost of the solar power. Photogalvanic cells (PGC) were studied immensely in 1980s as a cheap solar energy harvesting system. The photogalvanic  effect was first observed in 1925 by Rideal and Williams\citep{rideal_photogalvanic_1925}, and  it was  Rabinowitch that initially investigated the much studied iron-thionine  photogalvanic system\citep{rabinowitch_photogalvanic_1940}. Several other groups pursued the work both to understand the mechanism and to find the optimum device configuration for PGCs\citep{gomer_photogalvanic_1975,albery_foulds_photogalvanic_1979,sakata_photogalvanic_1977,suda_photogalvanic_1978,ferreira_photoredox_1977}. These studies together with the work on semiconductor electrochemistry by Gerischer\citep{gerischer_electrochemical_1966,gerischer_sensitization_1968}, Nozik\citep{nozik_photoelectrochemistry:_1978} and Gratzel\citep{gratzel_energy_1983} led the design of dye sensitized solar cells(DSSC)\citep{oregan_low-cost_1991}.  The analytical analysis of PGCs proposed by Albery and Archor \citep{albery_photogalvanic_1978} showed the possibility of high performance PGCs given certain conditions of device geometry and chemistry. Use of micelles in photogalvanic cell have been suggested by Groenen \emph{et al.} \citep{groenen_triton_1984} in order to increase dye solubility and suppress back-reaction.  Recently, a set of empirical studies examined a variety of dye/mediator couples,  including the study of several dyes by Gangorti \emph{et al.} \citep{gangotri_use_1996,gangotri_use_1997,gangotri_studies_2000,gangotri_studies_2010}, the effect of surfactants by Genwa \citep{genwa_photogalvanic_2008,genwa_comparative_2009} and even the use of mixed dyes by Lal \emph{et al.} \citep{lal_use_2007}. The highest efficiency of a PGC is claimed by Bhimwal and Gangotri to be 1.62\% with methyl orange as photosensitizer dye \citep{bhimwal_comparative_2011}, which expresses how far these devices are from practical use. Selective electrodes, fast electrode kinetics and high solubility of the dyes are the main unsatisfied properties of a good PGC. We propose a change in the configuration of the PGC that lightens up some of the hard to achieve requirements and we justify our proposed structure by analytical and numerical analysis.      

\begin{figure}[t]
\centering
\includegraphics[width=0.7\linewidth,natwidth=776,natheight=501]{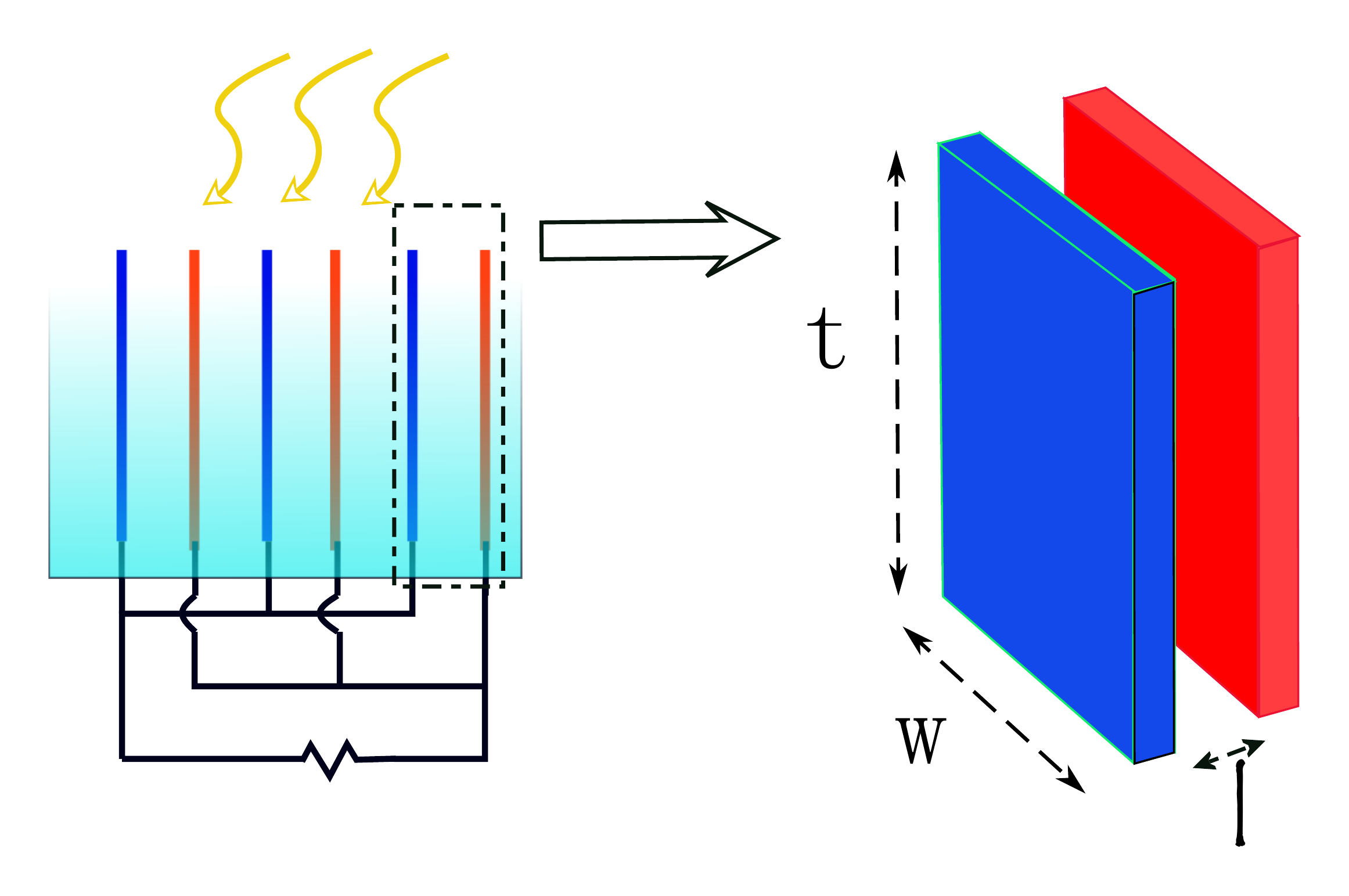}
\caption{Schematic of the vertical photogalvanic cell. The left figure shows several aligned vertical cells to cover a large area. Each cell consists of two parallel electrodes, with small spacing, $l$. Each electrode should be selective to one of the redox couples.}
\label{fig:PGScheme0}
\end{figure}

Instead of illuminating the device through the electrode as was done in the previous work, we suggest vertical alignment of  PGCs so that the light comes in from the gap between the two electrodes as shown in Figure~\ref{fig:PGScheme0}. This way, the light absorption and hence electron generation is spread through the depth of the device. As a result of larger absorption length, smaller current densities are expected and fast electrode kinetics are no longer required. The depth can be adjusted according to the concentration of the dyes, and thus deeper cells enable low solubility dyes to be employed. Multiple devices stack next to each other to cover surfaces. We suggest the investigation of this structure due to more relaxed requirements and higher possible efficiency.

 In the next section, the working principle of PGCs will be explored followed by the outlines of the theoretical work on conventional cells. The design guideline for efficient cell will be presented as was published by Albery et. al \citep{albery_foulds_photogalvanic_1979}.  The analysis of vertical cells is then presented and the requirements for high efficiencies will be derived from mathematical modeling. Several design factors will be discussed and compared to the conventional cell and the advantages of the new configuration are shown. In the simulation section the framework of a 2D computer model for PGCs is explained. Both cells are then modeled and optimized, assuming in one case known properties of dyes and mediators, and in the second case given dyes and mediators that should be physically realizable which gives a target configuration for PGCs. The expected efficiencies are compared. The benefits of the vertical cell are demonstrated and the target device parameters are shown to make the cell more viable than the conventional cell.

\section{Background}
\begin{figure}[!ht]
\centering
\includegraphics[width=0.7\linewidth,natwidth=450,natheight=300]{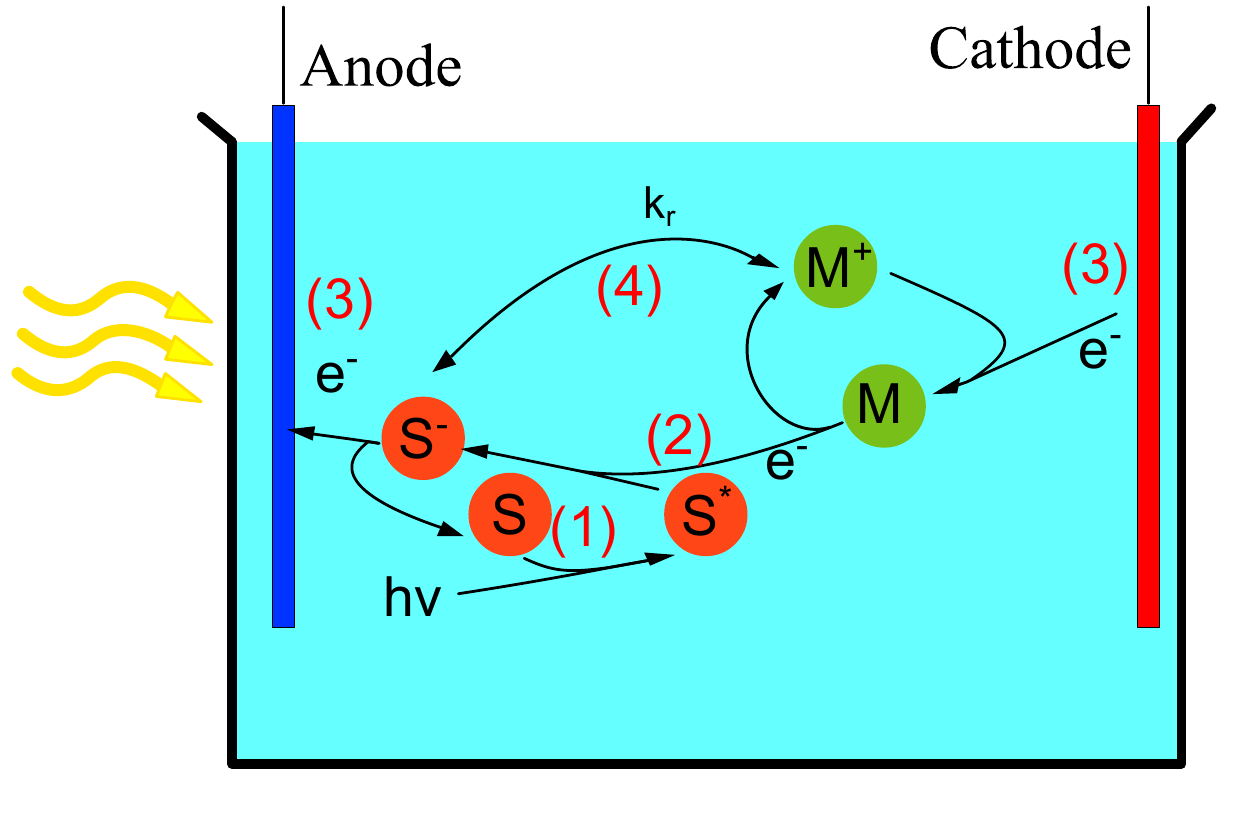}

\caption{Schematic of a traditional photogalvanic cell. The work flow is (1) light absorption and dye excitation.(2) Quenching of the excited dye and production of reduced dye and oxidized mediator.(3) Each redox couple interacts with one electrode and produces a current. (4) The bulk recombination tends to push the cell back to the equilibrium with rate constant $k_r$.}
\label{fig:PGScheme}
\end{figure}

The power generation process in photogalvanic cells starts with the photo-excitation of dissolved dye followed by dye reaction with an electrolyte redox system called  a mediator. The excited dye can be reduced or oxidized by the mediator depending on the dye and mediator combination selected. The redox couple and the stabilized dye then can react on the electrodes to generate current. In this work, the classic process of a dye-electron donor will be explored, whose schematic is shown in Figure \ref{fig:PGScheme}. 

The excitation step happens almost instantly after the absorption of the photon. The relaxation process happens at a rate of $k_{Relax}$ which is usually a very fast process and takes $10^{-12} \sim 10^{-9} s $ (reaction (\ref{eq:PGdyeExcitation}), in parentheses), 
\begin{eqnarray}
\label{eq:PGdyeExcitation} S+h\nu \rightarrow S^*(\xrightarrow{k_{Relax}} S).
\end{eqnarray} 
If the excited dye lives long enough to diffuse in the electrolyte and interact with a charge mediator, the excited dye will be quenched which results in two charged species;
\begin{eqnarray}
\label{eq:PGexcitation} S^*+M\xrightarrow{k_Q} S^{-}+ M^{+},
\end{eqnarray}  
where $S$, $S^*$ and $S^-$ are relaxed, excited and reduced states of the sensitizing dyes and $M/M^+$ is the mediating redox couple. Since the products are at high energy levels, they will recombine in the bulk with the rate constant  of $k_r$ which is the main loss mechanism of the cell,
\begin{equation}
S^{-}+ M^{+}\xrightarrow{k_r} S+ M.
\label{eq:PGrecombination}
\end{equation} 

In order to extract the absorbed energy, the products of reaction (\ref{eq:PGexcitation}) should diffuse to the electrodes before their recombination though reaction (\ref{eq:PGrecombination}). It is also desired that each  redox couple only interact with one of the electrodes in order to avoid electrode mediated recombination of the species \emph{i.e.} electrodes should behave selectively towards the couples. In this case,
\begin{align}
        \begin{array}{l}
\text{(Anode)    } S^{-}\rightarrow S +e^- \quad   \text{and}\\
\text{(Cathode)    } M^{+}+e^-\rightarrow M .
        \end{array}
        \label{eq:PGelectrodes}
\end{align} 

In the case of fast kinetics of these reactions, the electrode potentials will follow the potential of the redox couples, therefore, an open circuit potential difference of $\Delta E\approx|E_{S^-/S} -E_{M^+/M}|$ is expected from this cell. 

Albery examined the electrode-illuminated cell analytically and derived some design criteria for photogalvanic cells \cite{albery_photogalvanic_1978}. The differential equation governing the photo-absorption by dye molecules, the reaction of excited dyes and mediators, and the transport of species were solved simultaneously. The output of the analysis were four characteristic lengths of Table~\ref{tab:charLs}, which should be balanced according to cell requirement to achieve a high power efficiency.

\begin{table}[!ht]
    \centering
    \begin{tabular}{|l|l|p{9cm}|}
        \hline
        $X_l$ & $ l $ & Distance between electrodes \\ \hline
        $X_\varepsilon$ & $(\varepsilon[S])^{-1}$ & Light absorption length  \\ \hline
        $X_k$ & $({D}/{k_r[M^+]})^{1/2}$ & Typical diffusion length before recombination \\ \hline
        $X_g$&$({D}/{\phi_0 \varepsilon})^{1/2}$&Typical distance dye diffuses between photon absorption events\\ \hline
        
    \end{tabular}
    \caption{Characteristic lengths of photogalvanic devices. $ \varepsilon $ is the extinction coefficient of the dye. $[S]$ is the concentration of the light absorbing dye and $[M^+]$ is the concentration of the oxidized mediator. D is the diffusion coefficient. $ \phi_0 $ is the solar photon flux in units of [$mol \  m^{-2} s^{-1}$].}
    \label{tab:charLs}
\end{table}   

The complete absorption of the incoming light requires that the device be deep enough that most of the light be absorbed, or $X_\varepsilon<<X_l$. In order to extract the separated charge, generated $M^+$ ions need to travel to the illuminated electrode before their recombination, therefore, the distance over which generation is occurring should be less than the length over which it will likely diffuse before recombining through (\ref{eq:PGrecombination}), and hence $X_\varepsilon<<X_k$. Additionally, the excited dyes should be replaced by fresh ones before the arrival of the next photon in order to avoid solution bleaching, and hence, $X_\varepsilon<<X_g$.

The requirement of a net positive bulk generation requires that generation rate be faster than recombination rate therefore $X_g\leq X_k$. Finally, the maximum travel length is the distance between the electrodes, therefore the other three length constants should be smaller than $X_l$. The following formula was suggested for these values: 

\begin{equation}
10X_\varepsilon\approx X_g \approx \frac{1}{2}X_k < X_l.
\end{equation}  
These relations define the approximate conditions for the optimized cell. Using the typical values of $ D $, $ \varepsilon $ and $ I_0 $, a set of parameters were derived by Albery \emph{et al.}  \citep{albery_foulds_photogalvanic_1979} to make an efficient cell as shown in Table \ref{tab:OptimizedPG}.
\begin{table}[!ht]
    \centering
    \begin{tabular}{|l|l|}
        \hline
        $X_g=10\ \mu m$ & $D=10^{-5}\ cm^2 s^{-1}, \varepsilon=100\ mM^{-1}cm^{-1},I_0=1.6\cdot10^{-7} mol\ cm^{-2}s^{-1}$  \\ \hline
        $X_\varepsilon=1\ \mu m$ & $\Rightarrow [S]=0.1\ M $  \\ \hline
        $X_k=20\ \mu m$ & $\Rightarrow k[M^+]=2.5\ s^{-1}$  \\ \hline
        $X_l>20\ \mu m$& \\ \hline
       
    \end{tabular}
   \caption{Albery's recipe for the optimal cell  \citep{albery_foulds_photogalvanic_1979}.}
     \label{tab:OptimizedPG}
    
\end{table}  

One last requirement is that electrode kinetics be fast compared to the mass transport and recombination rates, and thus satisfy the following conditions:
\begin{subequations}
\begin{align}
k^0 & >> \frac{D}{X_\varepsilon} \quad   \text{and}\\
k^0 & > \frac{D}{X_k}.
\label{eq:AlberyOpk0}
\end{align}
\end{subequations}
The first condition ensures that the product species are generated close enough to the electrode to  be able to  interact with it and the second condition provides for a higher chance of electron extraction than bulk recombination.

As explained briefly in the introduction section, the latter condition of electrode kinetics is hard to satisfy, particularly in case of selective electrodes as any surface modification impedes the electron transfer between ions and the electrode. Additionally, fast electrode kinetics is incompatible with slow bulk reactions according to Marcus theory \citep{marcus_theory_1965}. In other words, no dye/mediator/electrode combination is likely to be found that offers both fast electrode kinetics and slow bulk recombination. Finally, the solubility of the dyes are much lower than the requirements of Table \ref{tab:OptimizedPG} \citep{albery_development_1982}, as a result, photogalvanic cells show poor efficiencies. Arranging the light path to be parallel to the electrode surfaces, as shown in Figure \ref{fig:PGScheme0}, is now shown to alleviate a number of constraints.  We investigate the requirements of target vertical cell in the next section and show that the proposed cell is less demanding in these two areas, i.e.,  moderate electrode kinetics  and dyes with low solubility can still be utilized in an efficient vertical cell.

\section{Analytical analysis of the vertical photogalvanic cell}
The vertical photogalvanic cell of Figure \ref{fig:PGScheme0} with the reactions shown in Figure \ref{fig:PGScheme}, is modeled in order to estimate the feasibility of the requirements for such a cell to work efficiently. First, an analytical model is presented that allows defining guidelines for design of efficient target vertical cells. This is followed by a numerical simulation, allowing the cell efficiency to be estimated.
 
 A comparison of maximum generation rate-which happens at the illuminated electrolyte surface- and the quenching rate constants, shows that even at small concentrations of mediator, the charge separation of reaction (\ref{eq:PGexcitation}) happens much faster than the initial photo-excitation of the dyes, reaction (\ref{eq:PGdyeExcitation}) (as has been previously assumed by Albery~\citep{albery_photogalvanic_1978}). Therefore, the two stages of light absorption and charge separation can be simplified into a single reaction of :
\begin{equation}
S+M\xrightleftharpoons[k_{r}]{G_{op}+k_f} S^-+M^+,
\end{equation}
where $ S/S^- $ represent the two states of the dye and $ M/M^+ $ those of the mediator. $G_{op}$, $k_f$ and $k_r$ are bulk reaction rates for optical generation, dark forward reaction and bulk recombination, respectively.   $ M/M^+ $ concentrations are assumed constant in the analysis by adding a condition that mediator concentration is much larger than that of the dye, following Albery~\citep{albery_photogalvanic_1978}, in order to enable an analytical solution. As a result of illumination to the gap, the optical generation, $ G_{op} $, varies through the depth.  This generation is averaged over the cell depth in our one dimensional analysis and therefore is independent of dye concentration as long as the cell is built deep enough to absorb all the incoming light. The assumptions of uniform generation and constant mediator concentration are removed in numerical analysis provided in the next section.


A diffusion-recombination reaction mechanism is considered for the transport in the bulk at steady-state,
\begin{equation}
D\frac{\pd^2 [S^-]}{\pd x^2}+G_{op}-k_r[S^-][M^+]=0. 
\label{eq:PGTATransport}
\end{equation}
In order to simplify the expressions, it is useful to rewrite the equations in dimensionless form. The length is normalized to the cell length, $l$, and the concentrations to the dark dye concentration, $[S_d]$. The unitless bulk reaction will be as follows,
\begin{align}
\frac{\pd^2 u}{\pd \chi^2}+&\alpha^2-\beta^2u=0 ,\quad \quad \text{where} \nonumber\\
\chi=x/l&,\quad u=[S^-]/[S_d],\quad \alpha^2=\frac{G_{op}\ l^2}{D\ [S_d]}\quad  and \quad \beta^2=\frac{l^2\ k_r\ [M^+]}{D}.
\label{eq:PGTDR}
\end{align}   
Parameter $\alpha$ compares the cell length to generation length, the distance dye diffuses before being hit by a photon. A large $ \alpha $ guarantees dye excitation before traveling the length of the cell. $\beta$ represents the cell length compared to recombination length, which is the diffusion distance of the charged states before recombination in the bulk. A small $ \beta $ is desired in order to increase the chance of charge extraction.   For the analytical analysis section,  completely selective electrodes with fast kinetics are assumed to be employed, where each electrode interacts only with one redox couple. The optimum cell performance conditions, which we are looking for, happen under this condition which guarantees minimum recombination. (The effect of the non-perfect mediator discrimination will be explored in the numerical analysis). This assumption allows us to assign all the current from the left electrode to the interaction with the $S/S^-$ couple, therefore,
\begin{align}
D\frac{d[S^-]}{dx}|_{x=0}&=\frac{J}{F} \quad 
\text{which translates to} \\
\label{eq:PGTBC1} \frac{d u}{d \chi}|_{\chi=0}&=m, \quad m=\frac{J\cdot l}{F D [S_d]},
\end{align}
where $ J $ is the cathodic current density, $ F $ is the Faraday constant, $ D $ is the diffusion coefficient and $ m $ is  the normalized current density.  
On the right electrode, no electron transfer happens with dyes due to the complete selectivity assumption. The boundary condition is then
\begin{equation}
\label{eq:PGTBC2} \frac{d u}{d \chi}|_{\chi=1}=0.
\end{equation}
Solving equation (\ref{eq:PGTDR}) with boundary conditions of (\ref{eq:PGTBC1}) and (\ref{eq:PGTBC2}) results in a normalized concentration profile as follows
\begin{equation}
u=\frac{\alpha^2-m \beta cosh(\beta -\beta \chi)csch(\beta)}{\beta^2}.
\label{eq:PGTSol}
\end{equation}
All device characteristics can be derived from equation \ref{eq:PGTSol}- most importantly, the current density, $ m $, which relates to $ u $ through (\ref{eq:PGTBC1}). We can define the cell efficiency in terms of concentration to be able to calculate cell parameters, $l$, $[S_d]$, $[M^+]$ and $k_r$, for the optimized cell. The concentration at  the surface of the electrode can be written as
\begin{equation}
\label{eq:PGTCurrent}u_0=u|_{\chi=0}=\frac{\alpha^2-m \beta coth(\beta)}{\beta^2}\Rightarrow m=\frac{\alpha^2-\beta^2u_0}{\beta  coth(\beta)}.
\end{equation} 
 
 The output voltage of the cell, the difference in electrode's electrochemical potential, is also normalized. The unitless potential difference, $\Delta P$ can be calculated as equation (\ref{eq:PGTPotenail})   
\begin{align}
P&=\frac{F}{RT}E\\
\Delta E&=E_2-E_1=E^0_{Y/Z}+\frac{RT}{F}ln\frac{[M^+]}{[M]}-E^0_{A/B}-\frac{RT}{F}ln\frac{[S]}{[S^-]}\\ 
\label{eq:PGTPotenail} [M^+],[M]&\approx constant \Rightarrow \Delta P=\Delta P^0 +ln(u_0^{-1}-1) \\
\Delta P^0&=\frac{F}{RT}(E^0_{M/M^+}-E^0_{S/S^-})
\end{align}

The efficiency of the cell can then be calculated by dividing the product of the output current and voltage by the incoming light power. Using equations (\ref{eq:PGTPotenail}) and (\ref{eq:PGTCurrent}), the efficiency can be written in the following form,
\begin{align}
\eta&=\frac{\left(t\cdot w \cdot J \right)\cdot \Delta E}{l\cdot w\cdot I_0 }\times 100 \%=\frac{FD[S_d]RT}{FI_0}\frac{t}{l^2}\ m\cdot \Delta P \times 100 \%\nonumber\\
\label{PGTEfficiency}&=\underset{\theta,\text{ max efficiency}} {\underbrace{\frac{D[S_d]RT\Delta P^0}{I_0}\frac{t}{l^2}\ \frac{\alpha^2}{\beta  \coth(\beta)}}} \underset{\psi,\ \text{load dependent}}{\underbrace{\left(1-\frac{\beta^2 u_0}{\alpha^2} \right)\left(1-\frac{ln(u_0^{-1}-1)}{\Delta P^0}\right)}} 100 \%.
\end{align}
 
The two right hand side terms are functions of $ u_0 $ and therefore $ m $, the current density, which is  dependent on the load connected to cell. In order to deliver the maximum efficiency, one should maximize this part, $ \psi $,  by adjusting the load and make those terms as close to unity as possible. The part that mainly governs the magnitude of the efficiency is the leftmost product in (\ref{PGTEfficiency}), called $\theta$, that needs to be maximized by adjusting the cell parameters. We first look for the cell conditions that optimize this term, then adjust load, and therefore $u_0$, to maximize the two RHS terms. 

\subsection{Geometry optimization }
Inserting the values of $\alpha$ and $\beta$ into $\theta$, one can write this efficiency term in the form of
\begin{align}
\theta&=\frac{D[S_d]RT\Delta P^0}{I_0}\frac{t}{l^2}\ \frac{\alpha^2}{\beta  coth(\beta)}=\frac{D[S_d]RT\Delta P^0}{I_0}\frac{t{G_{tot}}l^2/(tD[S_d])}{l^2 l\sqrt{\frac{k_r[M^+]}{D}}\coth{\left(l\sqrt{\frac{k_r[M^+]}{D}}\right)}}\nonumber\\
\label{eq:PGTtheta}&=\frac{RT\Delta P^0 G_{tot}}{I_0}
\frac{1}{l \sqrt{\frac{k_r[M^+]}{D}}\coth{\left(l\sqrt{\frac{k_r[M^+]}{D}}\right)}}.
\end{align}

\begin{figure}[!ht]
\centering
\includegraphics[width=0.7\linewidth,natwidth=640,natheight=480]{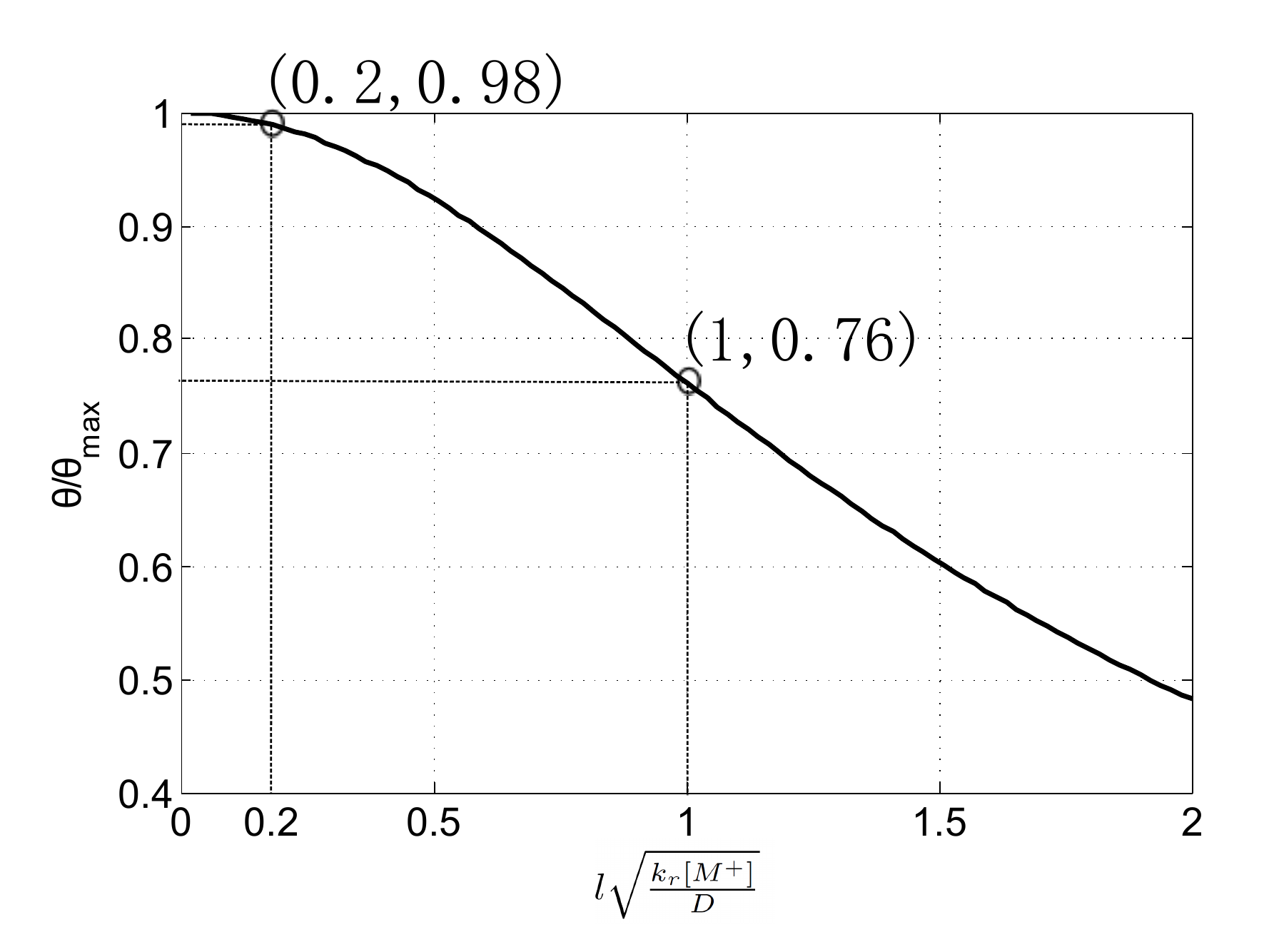}
\caption{Load independent efficiency vs $l\sqrt{\frac{k_r[M^+]}{D}}$. $ \theta_{max} $ happens at very small device lengths, however, the device length should be balanced between performance and fabrication limitation.  }

\label{fig:PGOpLength}
\end{figure}

All the variable parameters of equation (\ref{eq:PGTtheta}) are collected in the second term which represents a half-bell shaped function of $ \beta $ whose maximum occurs at zero, as depicted in Figure \ref{fig:PGOpLength}. Charge extraction at the electrodes always competes with bulk recombination, therefore a larger $ \beta $ (faster bulk recombination or wider device) consistently reduces the efficiency. Consequently, the cell length should be decreased to the extent that is allowed by the manufacturing limitations to have an efficient cell. One can see that $\theta$ still has 76\% of its maximum value when $l\sqrt{\frac{k_r[M^+]}{D}}\simeq 1$, which gives some room to deviate from the maximum point without a huge efficiency loss. 

 Assuming a slow - but feasible - bulk recombination rate of $ k_r=0.5\times 10^3\ M^{-1}s^{-1} $, a typical diffusion constant $ D=10^{-5}\ cm^{2}s^{-1} $, and a cell length of $ l=100\ \mu m $, the mediator concentration should be smaller than $ 200\ \mu M $ in order to enable reasonable efficiency. Dye concentration should be at least 2 times smaller than that of the mediator so that its change does not disturb the mediator concentration profile, therefore $ [S_d]\simeq 100\ \mu M $.

The depth of the cell has no direct influence on $ \theta $ output power as long as all the incoming light is absorbed in the cell, so the depth should be kept greater than $4\epsilon [S]$. The next term that should be maximized in the  $ \Delta P^0 G_{tot} $ product in the numerator. $\Delta P^0 $ reflects the difference in the electrochemical potentials of two redox couples which should ideally match the HOMO-LUMO levels  of the dye minus the required overpotential to drive the electron transfer. Therefore, the larger the difference is, the more significant the output voltage achievable. However, that would result a smaller portion in incident photons to be absorbed and the $G_{tot}$ to drop since lower energy photons cannot excite the dyes.  Balancing the trade-off between $G_{tot}$ and $\Delta P^0$ leads to an optimum HOMO-LUMO level difference of $1.4\ eV$, which in turn leads to $\theta_{max}=24\ \%$. This is the maximum efficiency regardless of the effect of the load, \emph{i.e.} $ I_{SC} \times V_{OC} /I_0$. Smaller separation between electrodes leads to higher efficiency - for example with $ 20\ \mu m $ separation, $ \theta_{max} $ can be expected to reach $ 31\ \%$ as the bulk recombination loss drops significantly.    

\subsection{Load optimization}
\begin{figure}[!ht]
\centering
\includegraphics[width=0.8\linewidth,natwidth=640,natheight=480]{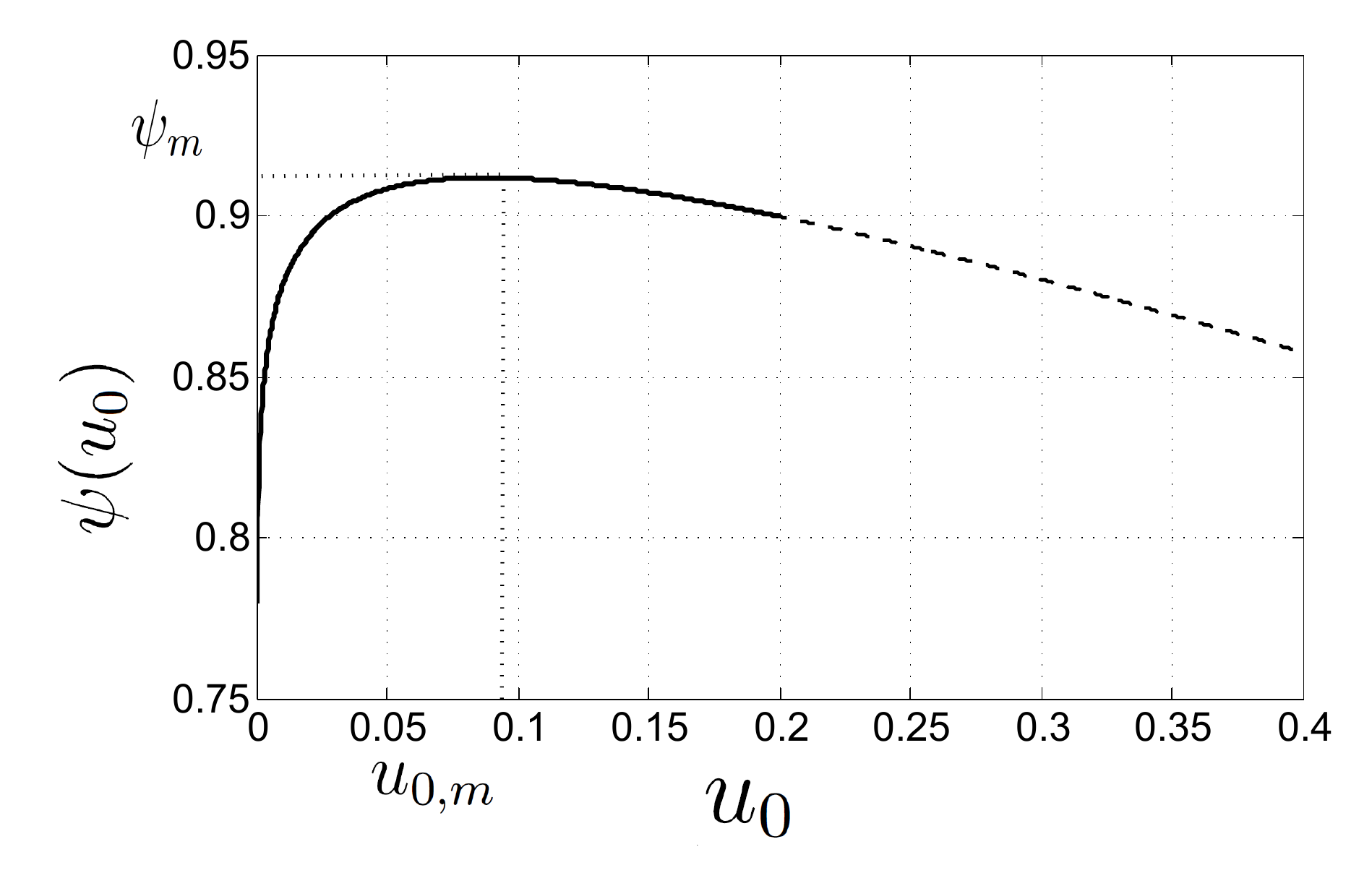}
\caption{the variation of $ \psi$ with respect to $ u_0 $. $ l=100\ \mu m $,$ [M_d] =200\ \mu M$, $ [S_d] =100\ \mu M$, $ k_r=0.5\times10^3\ M^{-1}s^{-1}$ and $ \Delta E= 1\ V $. }
\label{fig:psi}
\end{figure}

Both bracketed terms in the load dependent part of equation \ref{PGTEfficiency} , $\psi$, should approach unity in order to achieve maximum efficiency. For this section, this condition is assumed to be satisfied and an optimum load condition is calculated. The assumption is subsequently shown to be valid. Keeping that in mind, and neglecting the product of the small terms, $\psi$ can be approximated as : 
\begin{align}
\label{eq:PGTLoadApp1}\psi(u_0)&={\left(1-\frac{\beta^2 u_0}{\alpha^2} \right)\left(1-\frac{ln(u_0^{-1}-1)}{\Delta P^0}\right)}\simeq 1-\frac{\beta^2 u_0}{\alpha^2}-\frac{ln(u_0^{-1}-1)}{\Delta P^0}.
\end{align}
Differentiating with respect to $u_0$ shows a maximum at $u_{0,m}=\frac{\beta^2-\sqrt{\beta^2-4\alpha^2/\Delta P^0}}{2\beta^2} $ and $ \psi_m\simeq 0.91 $. Figure \ref{fig:psi} shows the variation of $ \psi$ with respect to $ u_0 $ for some typical cell parameters, where $ u_{0,m} $ is located very close to zero. The maximum efficiency load for this case happens close to the short circuit conditions which happens at  $ u_{0}=0 $ according to equation (\ref{eq:PGTCurrent}).   

\subsection{Estimation of maximum efficiency}
Altogether, total efficiency of the vertical photogalvanic cells would appear to go as high as $ 28 $ and $ 22\ \% $ for $20 $ and $100\ \mu m$ device lengths, respectively. The conditions for such performance are given above. In terms of device length constants, one can conclude the recombination length $ X_k$ should be larger than the device length $X_l $  as shown in Figure~\ref{fig:PGOpLength}. The light absorption in the vertical configuration is not limited to the electrode separation, however, since all the light needs to be captured, the cell thickness must be much larger than $ X_\varepsilon $, i.e. $ d>>X_\varepsilon $. As calculated above, the  $l\sqrt{\frac{k_r[M^+]}{D}}$ should be smaller than unity, which puts limits on bulk recombination rate, mediator concentration and device length. In practice not all the given device parameters are  readily achievable, but, as will be discussed later, they are more practical than those of traditional PGCs. In the next section these parameters will be fine tuned in a more realistic 2D system using electrodes with less than perfect selectivity.

\section{Simulation of the photogalvanic cell}

Numerical simulation enables the major assumptions of analytical analysis to be relaxed. A depth dependent generation, variable mediator concentration and imperfectly selective electrodes are modeled which gives a more accurate device analysis.  The photogalvanic cell was modeled in COMSOL Multiphysics software (COMSOL Inc., Palo Alto, CA) using a 2D geometry. COMSOL Multiphysics was chosen due to its ability to simultaneously solve several types of differential equations using the finite element method (FEM). The model includes two redox couples in the bulk and two selective electrodes with different reaction rates towards redox species. Bulk transport, generation-recombination and interaction on the electrodes were modeled using the \textit{Transport of Diluted Species} solver in COMSOL. The Butler-Volmer equation was solved on the electrodes with the built in \textit{ODE} solver and light absorption was modeled using the \textit{general PDE} solver, all of which were solved self-consistently in COMSOL. Both conventional and vertical photogalvanic cells were analyzed numerically first under optimum conditions and then with more readily available parameter values. The performance of the devices were compared for each scenario. 
\subsection{Equation set} 
Reactions (\ref{eq:PGdyeExcitation}), (\ref{eq:PGexcitation}), (\ref{eq:PGrecombination}) and (\ref{eq:PGelectrodes}) describe the interactions happening in  a working photogalvanic cells.  Below, the main bulk reaction is shown using both the notations used in the simulation and our analytical analysis,

\begin{equation}
\begin{cases}
 {o_1} + {r_2}\xrightleftharpoons[k_{r}]{G_{op}+k_f} {r_1} + {o_2}, \quad \text{or equivalently}\\
 S + M\xrightleftharpoons[k_{r}]{G_{op}+k_f} S^- + M^+.\\
 \end{cases}
\label{eq:PGMReaction}
\end{equation}
A reaction-diffusion system of equations is set up to model the transport of species in  the photogalvanic cell as shown in equation (\ref{eq:PGMTransport}),

\begin{equation}
\begin{cases}
 \frac{\partial c_i}{\partial t}+\nabla.(-D_i \nabla c_i)=R_i  \;\;\; i=o_1,r_1,o_2,r_2,\\
R_{o_1}=R_{r_2}=k_{r}[r_1][o_2]-k_{f}[o_1][r_2]-Gop,\\
R_{r_1}=R_{o_2}=-k_{r}[r_1][o_2]+k_{f}[o_1][r_2]+Gop.
 \end{cases}
\label{eq:PGMTransport}
\end{equation}

On the electrodes, reactions happen at different rates. The selectivity constraint requires that each electrode has fast kinetics with one couple and slow kinetics with the other one and the dominant reaction be that of equation (\ref{eq:PGMelectrodes}). The Butler-Volmer equation is used to describe the electron transfer at the interfaces,   
 \begin{flalign}     
 &\text{(On electrode 1)    } r_1\rightarrow o_1 +e^-,\nonumber\\
 \label{eq:PGMelectrodes}&\text{(On electrode 2)    } o_2+e^-\rightarrow r_2, \\
\label{eq:PGMBV}&J_/F=k_{1}^0\left([o_1]e^{-\alpha f \eta_1}-[r_1]e^{(1-\alpha) f \eta_1}\right)+k_{2}^0\left([o_2]e^{-\alpha f \eta_2}-[r_2]e^{(1-\alpha) f \eta_2}\right),    
 \end{flalign}
 where $\eta$, the overpotential,  is the difference between electrode's potential and the standard potential of the redox couple ($\eta_a=E_{elec}-E^0_a$). $ \alpha $ is the transfer coefficient and chosen to be 0.5 in this work which represents a symmetric energy barrier for electron transfer. $k_{1}^0  $ and $k_{2}^0  $ are standard rate constants of an electrode's interaction with redox couples \citep{bard_electrochemical_2001}.
%

The light absorption follows a Beer-Lambert law behavior in which the reduction in flux that happens in a layer with thickness $ dl $ containing species $i$ with concentration $ c_i $, is given by equation (\ref{eq:PGMlightabs}) where  $ \varepsilon $ is the molar absorption coefficient,
\begin{equation}
d\phi=-(\varepsilon ln 10) c_d \phi dl.
\label{eq:PGMlightabs}
\end{equation}
The $ c_d $ term of \ref{eq:PGMlightabs} is the dye concentration in the relaxed state i.e. [o1]. This links the two physics systems, light absorption and transport, together in the process of building a self consistent model. 

The other linking variable is the generation term in equation (\ref{eq:PGMTransport}). Assuming absorption to happen at a constant wavelength for simplicity, one can approximate the optical generation rate as 
\begin{equation}
Gop\simeq \frac{\phi_{abs}}{ \Delta z}\simeq \frac{d\phi}{dz}[\frac{mol}{m^3 s}],
\label{eq:PGMGop}
\end{equation}
where $ z $ can be the vertical or horizontal direction depending on the illumination direction. As can be seen in equations (\ref{eq:PGMlightabs}) and (\ref{eq:PGMGop}), the generation rate is proportional to the number of absorbed photons at each location which is non-linearly related to the concentration of light absorbers in the relaxed state. All of the equation system were implemented in COMSOL and the time dependent answer of the system was observed with initial conditions of $ [S]=[S_d] $ and $ [M]=[M^+] $. The steady state results were then extracted after the time dependent variables reach a plateau. These steady state values were taken as performance figures of the cell and compared for vertical and conventional cells in the next section.
\subsection{Results}
Photovoltaic devices traditionally use electrodes that lie in a plane that is ideally perpendicular to the direction of the incident light. One can extract most of the generated charges by putting the extracting electrode close to the absorbing section (junction). For the same reason, photogalvanic cells are illuminated through one transparent electrode while the other electrode is typically kept in the dark. This way, the peak generation happens very close to the collecting electrode. Despite Albery's initial analysis that showed  $18\%$ efficiency\citep{john_albery_photogalvanic_1978}, he concluded later that some practical restrictions limit the performance to 5\%\citep{albery_development_1982}. He derived conditions for this  'optimal' cell as shown in Table \ref{tab:OptimizedPG}. Simulations of the classic iron-thionine cell, and of Albery's 'optimal' cell integrated with a pair of selective electrodes were performed in our 2D model and the device optimization resulted in a power conversion efficiency of 0.45\% and 3.7\%. Thus it is not surprising that the best experimentally measured performance from the PGC is 1.62 \%~\citep{bhimwal_comparative_2011} (in which no significant selectivity is present). To understand the ultimate performance expected from these devices, the cell parameters were investigated again in the 2D model. The optimized traditional configuration cell, listed in Table~\ref{tab:SidePGSim}, showed an efficiency of 2.07\%. Only the extreme case of completely selective electrodes raise the efficiency to 3.7\%.  

\begin{table}[!ht]
    \centering
    \small
    \begin{tabular}{|l|p{2.5cm}|p{2.5cm}|p{2.5cm}|l|}
        \hline
         & Albery's Optimal (Perfect Selectivity) & Albery's Optimal (Partial Selectivity) \citep{albery_foulds_photogalvanic_1979}& Iron-Thionine\citep{albery_development_1982} &Explanation \\ \hline \hline
        
        $E^0_{S}$ &$ E^0_{S} $ &$ E^0_{S} $ & 0.462 V  & dye standard potential \\ \hline
        $E^0_{M}$ & $E^0_{S}$+ 1.4 eV &$E^0_{S}$+ 1.4 eV& 0.77 V & $M/M^+$ standard potential   \\ \hline
        $E_{ph}$ & 1.6 eV & 1.6 eV & 2.07 eV  & dye band gap  \\ \hline
        $k_{L,1}$ & $ 10^{-3} m/s $ &$ 10^{-3} m/s $& $ 10^{-5} m/s $& rate constant , fast redox \\ \hline
        $k_{L,2}$ & 0 m/s&$ 10^{-10} m/s $& $ 10^{-12} m/s $& rate constant ,slow redox\\ \hline
        $k_{r}$ & $ 5 \times 10^2 M^{-1}s^{-1} $ &$ 5 \times 10^2 M^{-1}s^{-1} $&$ 5 \times 10^2 M^{-1}s^{-1} $ & bulk recombination rate \\ \hline
        $l$ & 50$\mu m$ & 50$\mu m$& 50$\mu m$  & cell length \\ \hline
        $[S_d]$ & 100 $mM$ & 100 $mM$& 100 $mM$  & dye concentration \\ \hline
        $I_0$ & $1000 W/m^2$ & $1000 W/m^2$ & $1000 W/m^2$ & light intensity \\ \hline \hline
        $\eta$ & 3.7 \%  & 2.07 \%& 0.45 \%& Efficiency \\ \hline
    \end{tabular}
    \caption{Device parameters and performance of electrode illuminated  photogalvanic devices. Efficiencies, $ \eta $, are as computed using COMSOL.}
    \label{tab:SidePGSim}
\end{table}

 It should be noted that even this low efficiency performance is impractical in reality. Some characteristics used in Table \ref{tab:SidePGSim} to compute Albery's 'optimal' are incompatible with each other and some are simply hard to achieve. For example, very few redox couples and electrodes meet the very fast electrode kinetics requirement of equation \ref{eq:AlberyOpk0}. Fast electrode kinetics, needed to produce high currents, are also incompatible with slow bulk reactions (needed to reduce recombination losses between mediators) according to Marcus theory \citep{marcus_theory_1965}. The actual solubility of the dyes are much lower than those assumed here, which leads to poorer performances in practice compared to the theory.

In the suggested vertical configuration of Figure \ref{fig:PGScheme0}, light absorption and charge extraction lengths have been decoupled, therefore a smaller dye concentration can be utilized to reduce the current density through the electrodes while not affecting the generated current per illuminated surface. Electrode kinetics need not to be particularly high if dye concentrations are low, and similarly diffusion lengths to electrodes can be relatively long (provided they are similar to or shorter than the recombination length). The selectivity level of each electrode - the difference in reaction rates towards the two redox couples - is investigated. The results show that a 6 to 7 order of magnitude difference in rate constants is enough to achieve an efficient cell. 

The vertical cell is modeled in COMSOL, and shows improvement in performance. The efficiencies of the target and iron-thionine cells were found to be $12.9\%$ and $6\%$, respectively for a $100\ \mu m$ cell length. Because of the partial selectivity of the electrodes and the concentration dependent optical generation that were neglected in the theoretical model, these values, achieved with parameters of Table \ref{tab:VerticalPGSim}, are smaller than the prediction of the analytical analysis. The parameters listed in Table~\ref{tab:VerticalPGSim} are chosen with physical feasibility in mind. It is desired to have as small a bulk recombination rate as possible, the suggested dye-mediator couple of the target case is assumed to have a recombination rate in the same range as the iron-thionine couple, which is one of the slower known bulk reaction rates. The value for iron-thionine couple is determined from literature to be $5\times 10^2 M^{-1}s^{-1}$ \citep{albery_development_1982}. The dye concentrations are limited to sub-millimolar range and electrode kinetic rate is on the order of $ 10^{-5} m/s $.

\begin{figure}[!ht]
\centering
\subfigure[I-V Characteristics]{
\includegraphics[width=.4\columnwidth,natwidth=585,natheight=450]{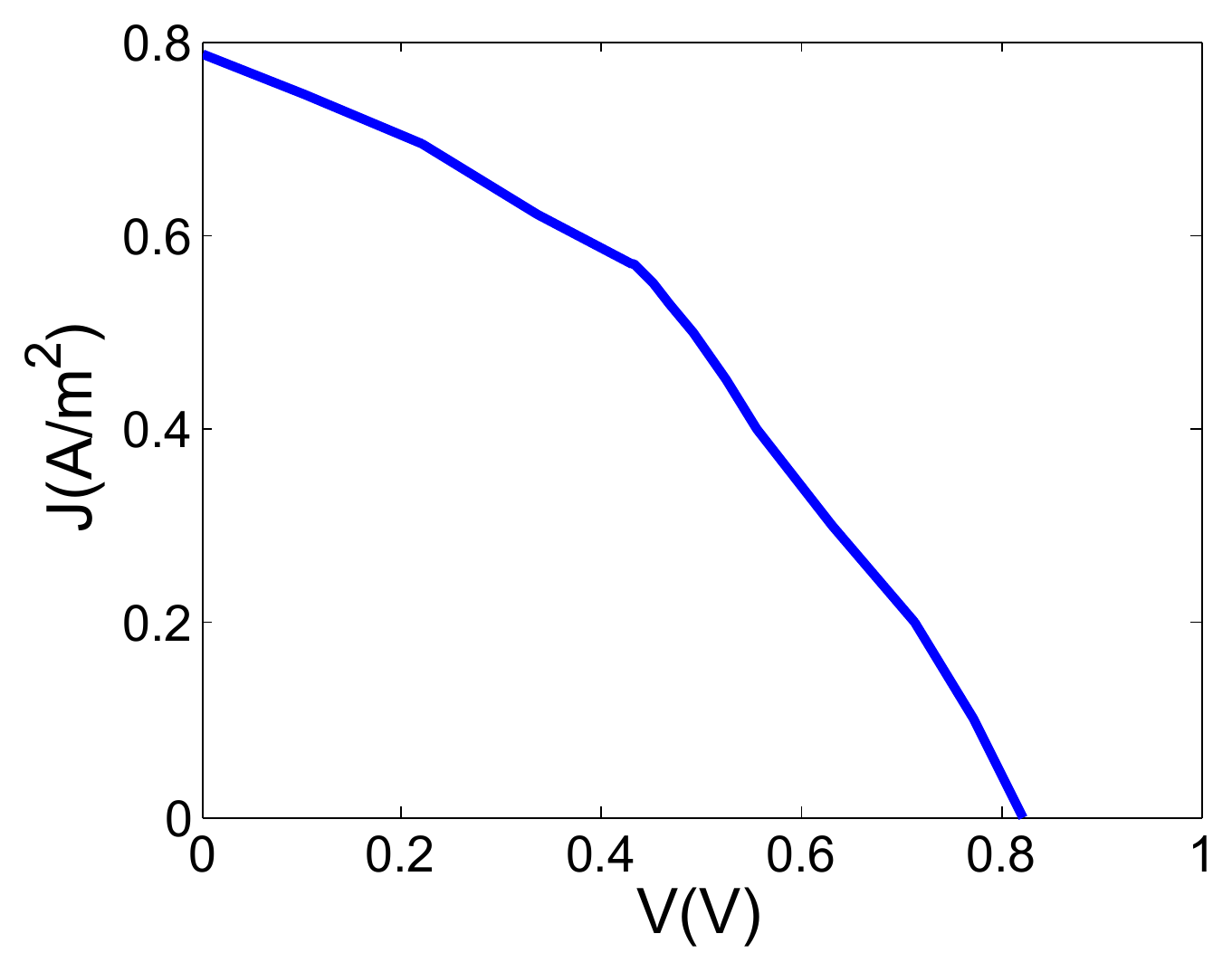}
\label{fig:CIV}
}
\subfigure[Cell Efficiency]{
\includegraphics[width=.4\columnwidth,natwidth=585,natheight=450]{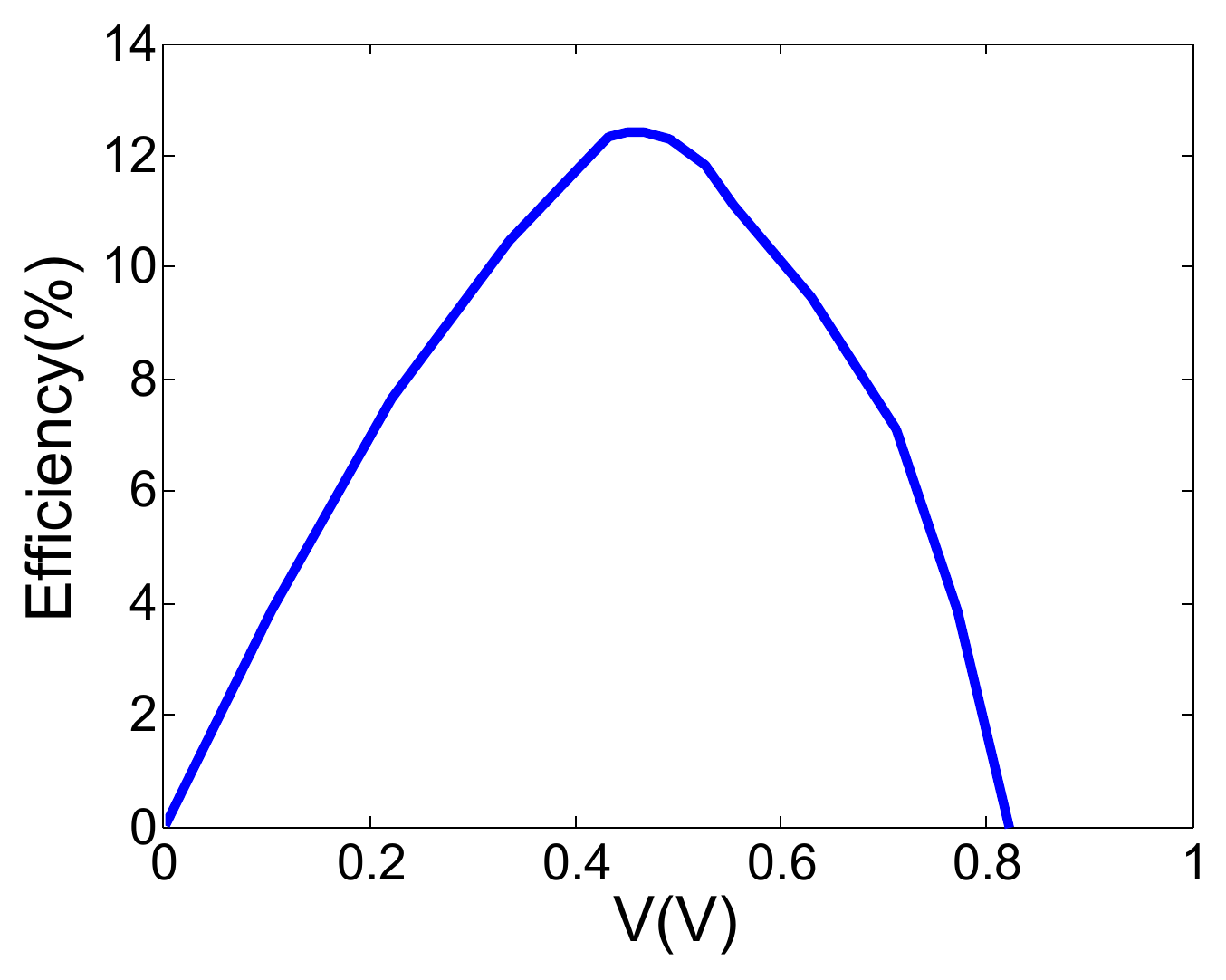}
\label{fig:CPV}
}

\subfigure[Light intensity at $ \eta_{max} $]{

\includegraphics[width=.4\columnwidth,natwidth=800,natheight=600]{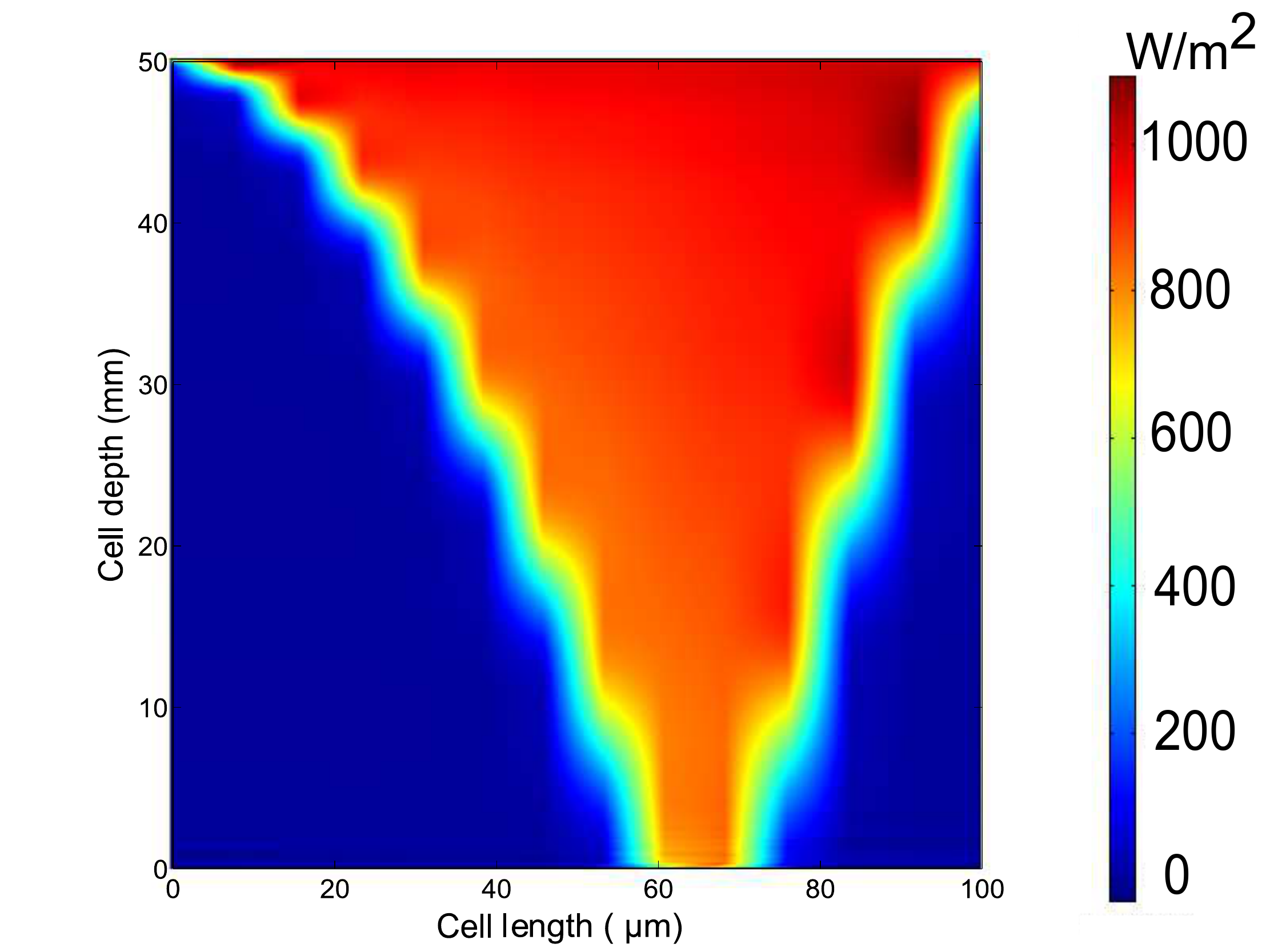}
\label{fig:CI0571}
}
\subfigure[$ D^- $ concentration at $ \eta_{max} $]{
\includegraphics[width=.4\columnwidth,natwidth=800,natheight=600]{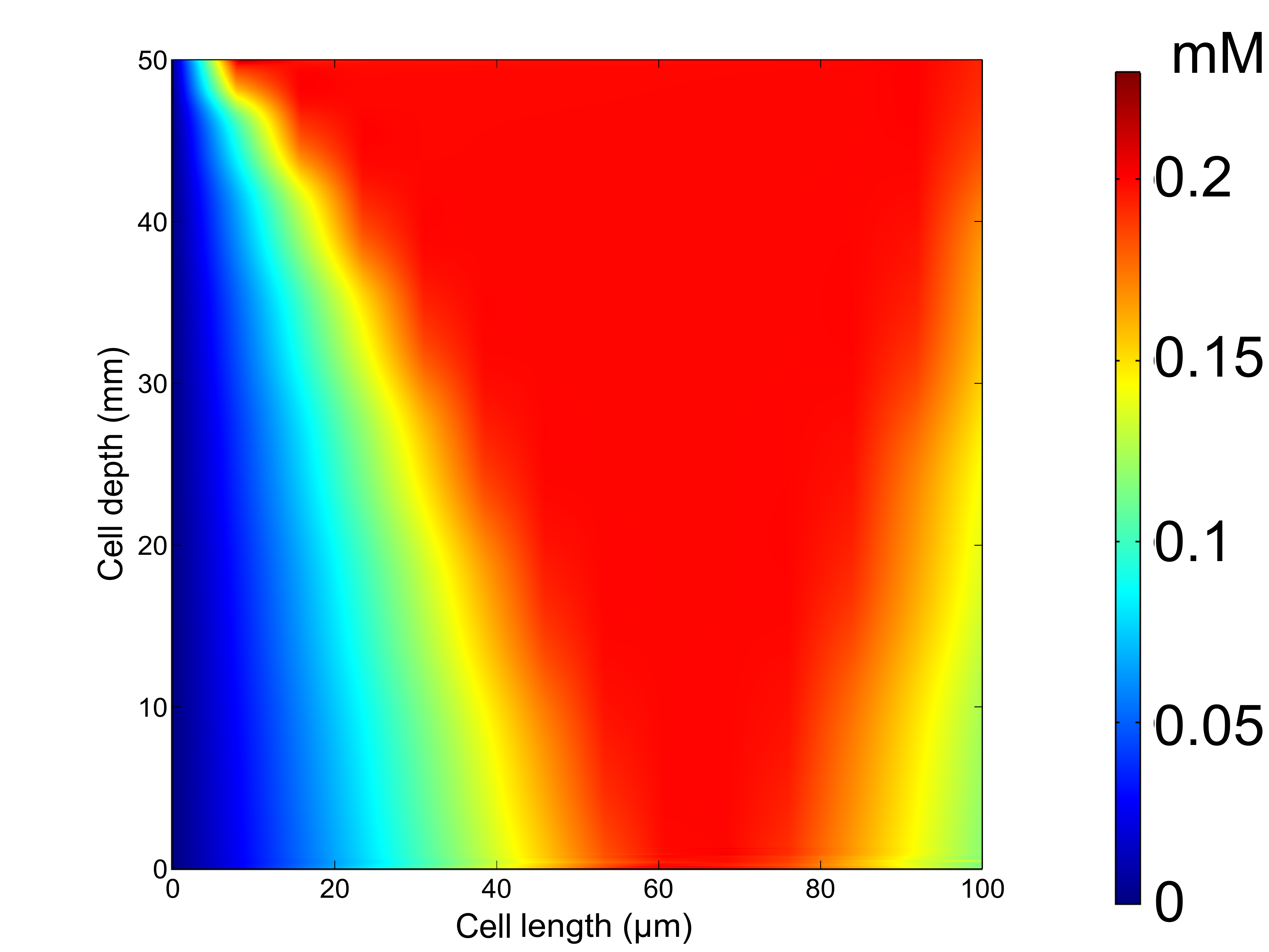}
\label{fig:CB0571}
}
\caption{Simulation results of the target vertical PGC. $ l=100 \mu m $,$ [M_d] =500 \mu M$, $ [S_d] =200 \mu M$, $ k_r=0.5\times10^3 M^{-1}s^{-1}$ and $ \Delta E= 1.4 V$  }
\label{fig:Comsol results}

\end{figure}

 Device characteristics of the target vertical cell are shown in Figures \ref{fig:CIV} and \ref{fig:CPV}.  The low fill factor observed in the Current-Voltage characteristics is mainly due to bulk and electrode recombination losses. Dye concentration and cell depth were swept in value to find the optimum concentration and size which enables both the use of the slow kinetics electrode (by reducing the generation rate) and the full absorption of the incoming light. The light intensity, $ I $, is plotted through the depth of the cell in Figure \ref{fig:CI0571}. The horizontal locations where $I$ goes to zero represents full absorption by the dye. In this graph there is small part where not all light is absorbed. The optimum dye concentration is $200\ \mu M  $, which is higher than the optimal theoretical value of the last section, because the effect of bleaching on the surface that was previously ignored in the theoretical analysis. Despite the bleaching, most of the incoming light is absorbed by over a depth of $50\  mm$. 

A comparison of the cell parameters in two configurations reveals the vertical cell to be less demanding. The required electrode kinetics is reduced by two orders of magnitude in the vertical structure and partial selectivity of seven order of magnitudes proves to be sufficient for efficient device performance. This is another advantage of the vertical cell since the $ 3.7 \% $ ultimate efficiency of the conventional cell was calculated based on complete selectivity and it drops to $ 2.07\ \%$   at a selectivity of $ 7 $ orders of magnitude. The concentration of the reduced dye, $ [S^-] $, is plotted in \ref{fig:CB0571}. It reaches zero on the dye-interacting electrode at maximum efficiency operation point, which agrees well with the theoretical prediction that the optimum load happens at very small value of $ u_{0,m} $, equation \ref{eq:PGTLoadApp1}.

\begin{figure}[!ht]
\centering
\subfigure[Top illuminated cell]{
\includegraphics[width=.4\columnwidth,natwidth=524,natheight=394]{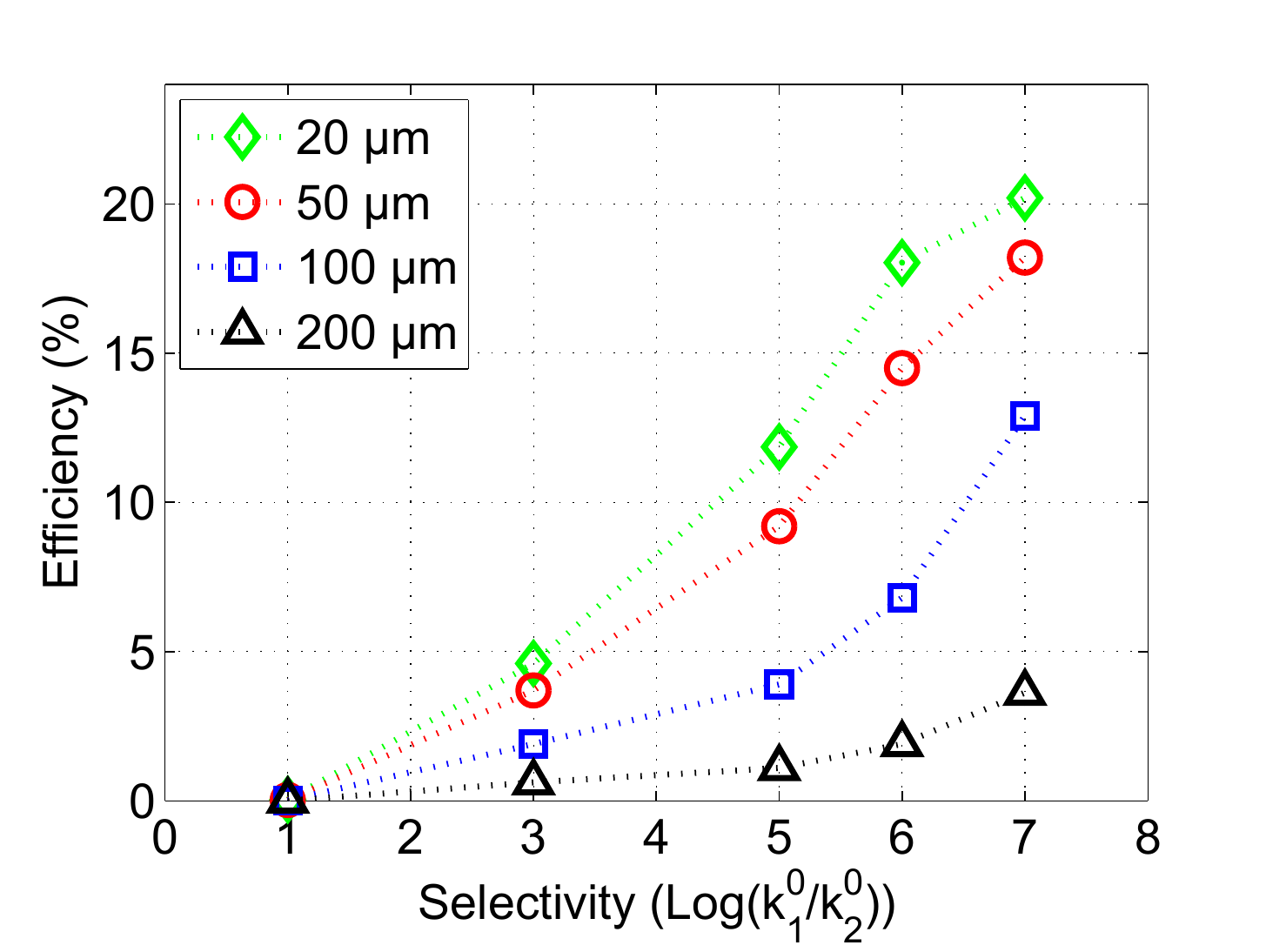}
\label{fig:etavsselectivityTop}
}
\subfigure[Side illuminated cell]{
\includegraphics[width=.4\columnwidth,natwidth=524,natheight=394]{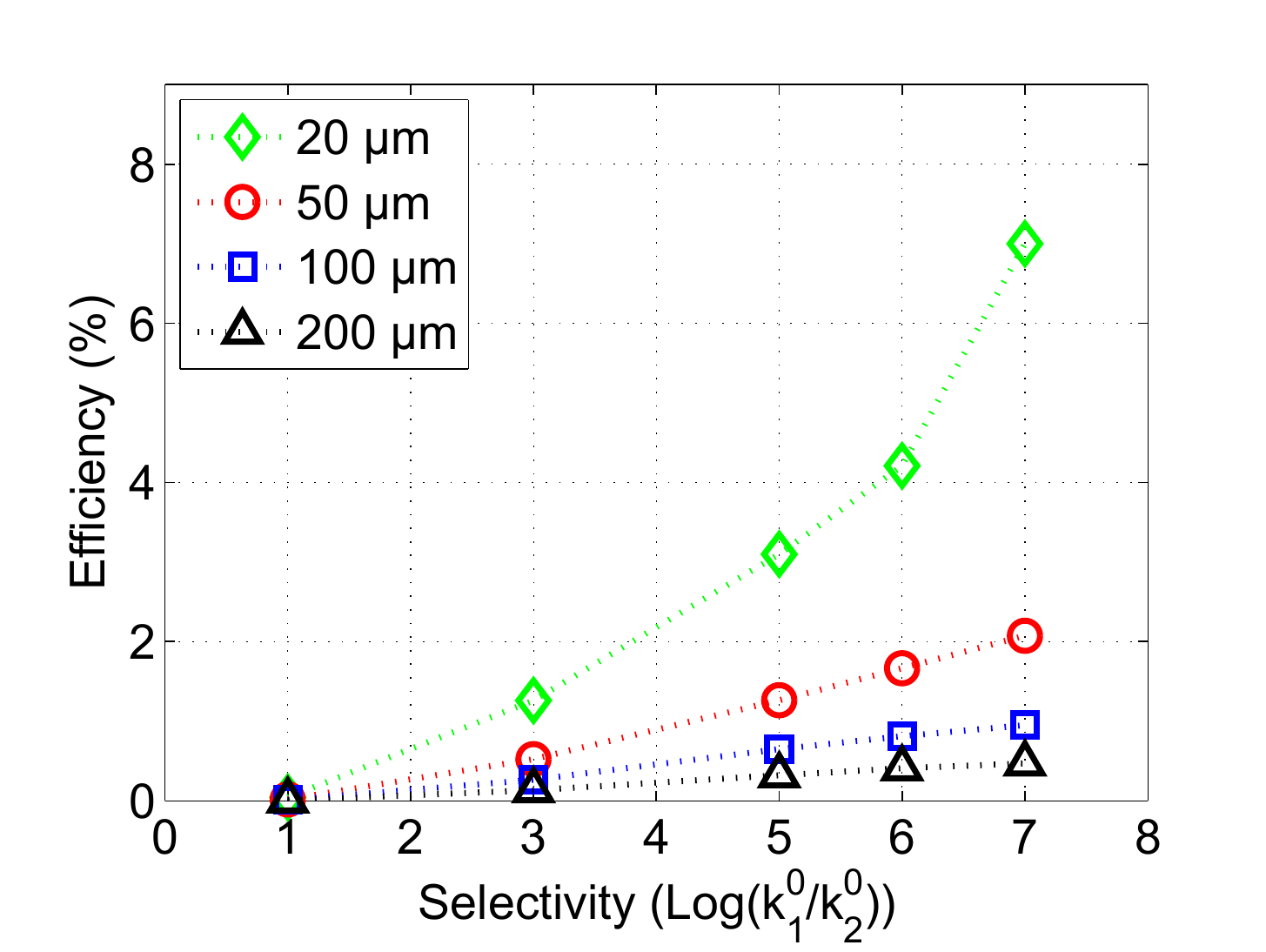}
\label{fig:etavsselectivitySide}
}

\caption{Effect of device length and electrode selectivity on cell efficiency for both top and side illuminated cells.}
\label{fig:etavsselectivity}

\end{figure}

As explained in the analytical section, electrode separation is inversely related to the performance. Figure~\ref{fig:etavsselectivity} shows the efficiency of top and side illuminated cells for different cell lengths and electrode selectivities. Device performance is more sensitive to selectivity in vertical cell for device lengths larger than $ 20\  \mu m $. Therefore, a pair of selective electrodes is crucial in making a practical vertical PGC. For the target cell, this length was set to $ 100\  \mu m $ for practical reasons, however, as shown in Figure~\ref{fig:etavsselectivity}, efficiencies up to 20.2 \% is achievable with thiner vertical cells. It can be seen that 12.9 \% efficiency of the $ 100\  \mu m $ vertical PGC is not achievable with any conventional PGC regardless of the geometry. 

Overall, promising device performance is expected with physically feasible parameters which are close to the maximum we think can be achievable. Further research is required to find dye/mediator couples in order to improve these parameters. Due to the small length of each cell, multiple cells must be fabricated in series to cover large areas.
\begin{table}[ht]
    \centering
    \small
    \begin{tabular}{|l|p{2.5cm}|p{2.6cm}|l|}
        \hline
         & Target vertical cell& Iron-Thionine vertical cell& Explanation \\ \hline \hline
        
        $E^0_{S}$ & $E^0_{S}$& 0.462 V& dye standard potential \\ \hline
        $E^0_{M}$ &$E^0_{S}$+ 1.4 eV &0.77 V & $M/M^+$ standard potential   \\ \hline
        $E_{ph}$ &1.6 eV & 2.07 eV & Dye band gap  \\ \hline
        $k_{L,1}$ & $ 10^{-5} m/s $& $ 10^{-5} m/s $& rate constant, fast redox \\ \hline
        $k_{L,2}$ &$ 10^{-10} m/s $ &$ 10^{-10} m/s $ & rate constant, slow redox\\ \hline
        $k_{r}$ &$  5\times 10^2 M^{-1}s^{-1} $ &$5\times 10^2 M^{-1}s^{-1} $ & bulk recombination rate \\ \hline
        $l$ & 100$\mu m$ &100$\mu m$ & cell length \\ \hline
        $[S_d]$ & 200$\mu M$ & 200$\mu M$  & dye concentration \\ \hline
        $t$ & 50 $mm$ & 50 $mm$  & cell depth \\ \hline
		$I_0$ &$1000 W/m^2$ &$1000 W/m^2$ & light intensity \\ \hline \hline
        $\eta$ & 12.9 \%&6 \%& Efficiency \\ \hline
    \end{tabular}
    \caption{Device parameters and performance of vertical photogalvanic devices. Efficiencies, $ \eta $, are as computed using COMSOL.}
    \label{tab:VerticalPGSim}
\end{table}

\section{Conclusion}
Vertical configuration photogalvanic cells are suggested and modeled.  The analysis of an individual cell shows this configuration should result in higher efficiencies than where the illumination is through the electrode. To be effective, sub-millimolar dye concentration and slow bulk recombination rate on the order of $10^3\ M^{-1}s^{-1}$ are required, which is not easy to achieve but not impossible as iron-thionine recombination rate is half this number. Electrode kinetics should be reasonably fast, but extending the light absorption through a depth of the cell makes  moderate electron transfer rate constants of $10^{-5}\ ms^{-1}$ sufficient, which is 100 times slower than the requirement for the traditional cell. The electrodes were assumed  to be completely selective in the analytical section, but optimizing this parameter in the numerical analysis revealed a need for 6 to 7 orders of magnitude difference in rate constants. So far, selectivity up to 3 orders was shown by our group\citep{usgaocar_semiconductors_2013}. Further work needs to be done in this area. All in all, we believe vertical PGCs can be used as cheap, low maintenance solar cells assuming that proper electrode-dye-mediator-electrode combination is found.   
\section{Acknowledgments}
The authors gratefully  acknowledge the financial support through Discovery and Strategic Grants from the Natural Sciences and Engineering Research Council(NSERC) of Canada and The Peter Wall Institute for Advanced Studies at UBC.     

\bibliographystyle{unsrt}
\bibliography{VPGAnalysis}

\begin{thebibliography}{27}
\expandafter\ifx\csname natexlab\endcsname\relax\def\natexlab#1{#1}\fi
\providecommand{\url}[1]{\texttt{#1}}
\providecommand{\href}[2]{#2}
\providecommand{\path}[1]{#1}
\providecommand{\DOIprefix}{doi:}
\providecommand{\ArXivprefix}{arXiv:}
\providecommand{\URLprefix}{URL: }
\providecommand{\Pubmedprefix}{pmid:}
\providecommand{\doi}[1]{\href{http://dx.doi.org/#1}{\path{#1}}}
\providecommand{\Pubmed}[1]{\href{pmid:#1}{\path{#1}}}
\providecommand{\bibinfo}[2]{#2}
\ifx\xfnm\relax \def\xfnm[#1]{\unskip,\space#1}\fi
\bibitem[{Rideal and Williams(1925)}]{rideal_photogalvanic_1925}
\bibinfo{author}{E.~K. Rideal}, \bibinfo{author}{D.~C. Williams},
\newblock \bibinfo{title}{Photogalvanic effect},
\newblock \bibinfo{journal}{J. Chem. Soc} \bibinfo{volume}{127}
  (\bibinfo{year}{1925}) \bibinfo{pages}{258}.
\bibitem[{Rabinowitch(1940)}]{rabinowitch_photogalvanic_1940}
\bibinfo{author}{E.~Rabinowitch},
\newblock \bibinfo{title}{The photogalvanic effect i. the photochemical
  properties of the thionine-iron system},
\newblock \bibinfo{journal}{The Journal of Chemical Physics}
  \bibinfo{volume}{8} (\bibinfo{year}{1940}) \bibinfo{pages}{551}. \URLprefix
  \url{http://link.aip.org/link/?JCPSA6/8/551/1}.
\bibitem[{Gomer(1975)}]{gomer_photogalvanic_1975}
\bibinfo{author}{R.~Gomer},
\newblock \bibinfo{title}{Photogalvanic cells},
\newblock \bibinfo{journal}{Electrochimica Acta} \bibinfo{volume}{20}
  (\bibinfo{year}{1975}) \bibinfo{pages}{13--20}. \URLprefix
  \url{http://www.sciencedirect.com/science/article/pii/0013468675850389}.
  \DOIprefix\doi{16/0013-4686(75)85038-9}.
\bibitem[{John~Albery and Foulds(1979)}]{albery_foulds_photogalvanic_1979}
\bibinfo{author}{W.~John~Albery}, \bibinfo{author}{A.~W. Foulds},
\newblock \bibinfo{title}{Photogalvanic cells},
\newblock \bibinfo{journal}{Journal of Photochemistry} \bibinfo{volume}{10}
  (\bibinfo{year}{1979}) \bibinfo{pages}{41--57}. \URLprefix
  \url{http://www.sciencedirect.com/science/article/pii/0047267079800362}.
  \DOIprefix\doi{10.1016/0047-2670(79)80036-2}.
\bibitem[{Sakata et~al.(1977)Sakata, Suda, Tanaka, and
  Tsubomura}]{sakata_photogalvanic_1977}
\bibinfo{author}{T.~Sakata}, \bibinfo{author}{Y.~Suda},
  \bibinfo{author}{J.~Tanaka}, \bibinfo{author}{H.~Tsubomura},
\newblock \bibinfo{title}{Photogalvanic effect in the thionine-iron system},
\newblock \bibinfo{journal}{The Journal of Physical Chemistry}
  \bibinfo{volume}{81} (\bibinfo{year}{1977}) \bibinfo{pages}{537--542}.
  \URLprefix \url{http://dx.doi.org/10.1021/j100521a009}.
  \DOIprefix\doi{10.1021/j100521a009}.
\bibitem[{Suda et~al.(1978)Suda, Shimoura, Sakata, and
  Tsubomura}]{suda_photogalvanic_1978}
\bibinfo{author}{Y.~Suda}, \bibinfo{author}{Y.~Shimoura},
  \bibinfo{author}{T.~Sakata}, \bibinfo{author}{H.~Tsubomura},
\newblock \bibinfo{title}{Photogalvanic effect in the thionine-iron system at
  semiconductor electrodes},
\newblock \bibinfo{journal}{The Journal of Physical Chemistry}
  \bibinfo{volume}{82} (\bibinfo{year}{1978}) \bibinfo{pages}{268--271}.
  \URLprefix \url{http://dx.doi.org/10.1021/j100492a003}.
  \DOIprefix\doi{10.1021/j100492a003}.
\bibitem[{Ferreira and Harriman(1977)}]{ferreira_photoredox_1977}
\bibinfo{author}{M.~I.~C. Ferreira}, \bibinfo{author}{A.~Harriman},
\newblock \bibinfo{title}{Photoredox reactions of thionine},
\newblock \bibinfo{journal}{Journal of the Chemical Society, Faraday
  Transactions 1: Physical Chemistry in Condensed Phases} \bibinfo{volume}{73}
  (\bibinfo{year}{1977}) \bibinfo{pages}{1085--1092}. \URLprefix
  \url{http://pubs.rsc.org/en/content/articlelanding/1977/f1/f19777301085}.
  \DOIprefix\doi{10.1039/F19777301085}.
\bibitem[{Gerischer(1966)}]{gerischer_electrochemical_1966}
\bibinfo{author}{H.~Gerischer},
\newblock \bibinfo{title}{Electrochemical behavior of semiconductors under
  illumination},
\newblock \bibinfo{journal}{Journal of The Electrochemical Society}
  \bibinfo{volume}{113} (\bibinfo{year}{1966}) \bibinfo{pages}{1174--1182}.
  \URLprefix \url{http://jes.ecsdl.org/content/113/11/1174}.
  \DOIprefix\doi{10.1149/1.2423779}.
\bibitem[{Gerischer et~al.(1968)Gerischer, Michel-Beyerle, Rebentrost, and
  Tributsch}]{gerischer_sensitization_1968}
\bibinfo{author}{H.~Gerischer}, \bibinfo{author}{M.~Michel-Beyerle},
  \bibinfo{author}{F.~Rebentrost}, \bibinfo{author}{H.~Tributsch},
\newblock \bibinfo{title}{Sensitization of charge injection into semiconductors
  with large band gap},
\newblock \bibinfo{journal}{Electrochimica Acta} \bibinfo{volume}{13}
  (\bibinfo{year}{1968}) \bibinfo{pages}{1509--1515}. \URLprefix
  \url{http://www.sciencedirect.com/science/article/pii/0013468668800763}.
  \DOIprefix\doi{16/0013-4686(68)80076-3}.
\bibitem[{Nozik(1978)}]{nozik_photoelectrochemistry:_1978}
\bibinfo{author}{A.~J. Nozik},
\newblock \bibinfo{title}{Photoelectrochemistry: Applications to solar energy
  conversion},
\newblock \bibinfo{journal}{Annual Review of Physical Chemistry}
  \bibinfo{volume}{29} (\bibinfo{year}{1978}) \bibinfo{pages}{189--222}.
  \URLprefix \url{zotero://attachment/598/}.
  \DOIprefix\doi{10.1146/annurev.pc.29.100178.001201}.
\bibitem[{Gratzel(1983)}]{gratzel_energy_1983}
\bibinfo{author}{M.~Gratzel}, \bibinfo{title}{Energy Resources through
  Photochemistry and Catalysis}, \bibinfo{publisher}{Elsevier},
  \bibinfo{year}{1983}.
\bibitem[{{O'Regan} and Gratzel(1991)}]{oregan_low-cost_1991}
\bibinfo{author}{B.~{O'Regan}}, \bibinfo{author}{M.~Gratzel},
\newblock \bibinfo{title}{A low-cost, high-efficiency solar cell based on
  dye-sensitized colloidal {TiO2} films},
\newblock \bibinfo{journal}{Nature} \bibinfo{volume}{353}
  (\bibinfo{year}{1991}) \bibinfo{pages}{737--740}. \URLprefix
  \url{http://dx.doi.org/10.1038/353737a0}. \DOIprefix\doi{10.1038/353737a0}.
\bibitem[{Albery and Archer(1978)}]{albery_photogalvanic_1978}
\bibinfo{author}{W.~J. Albery}, \bibinfo{author}{M.~D. Archer},
\newblock \bibinfo{title}{Photogalvanic cells: Part 3. the maximum power
  obtainable from a thin layer photogalvanic concentration cell with identical
  electrodes},
\newblock \bibinfo{journal}{Journal of Electroanalytical Chemistry and
  Interfacial Electrochemistry} \bibinfo{volume}{86} (\bibinfo{year}{1978})
  \bibinfo{pages}{1--18}. \URLprefix
  \url{http://www.sciencedirect.com/science/article/pii/S0022072878803519}.
  \DOIprefix\doi{10.1016/S0022-0728(78)80351-9}.
\bibitem[{Groenen et~al.(1984)Groenen, De~Groot, De~Ruiter, and
  De~Wit}]{groenen_triton_1984}
\bibinfo{author}{E.~J.~J. Groenen}, \bibinfo{author}{M.~S. De~Groot},
  \bibinfo{author}{R.~De~Ruiter}, \bibinfo{author}{N.~De~Wit},
\newblock \bibinfo{title}{Triton x-100 micelles in the ferrous/thionine
  photogalvanic cell},
\newblock \bibinfo{journal}{The Journal of Physical Chemistry}
  \bibinfo{volume}{88} (\bibinfo{year}{1984}) \bibinfo{pages}{1449--1454}.
  \URLprefix \url{http://dx.doi.org/10.1021/j150651a043}.
  \DOIprefix\doi{10.1021/j150651a043}.
\bibitem[{Gangotri et~al.(1996)Gangotri, Regar, Lal, Kalla, Genwa, and
  Meena}]{gangotri_use_1996}
\bibinfo{author}{K.~Gangotri}, \bibinfo{author}{O.~P. Regar},
  \bibinfo{author}{C.~Lal}, \bibinfo{author}{P.~Kalla}, \bibinfo{author}{K.~R.
  Genwa}, \bibinfo{author}{R.~Meena},
\newblock \bibinfo{title}{Use of tergitol-7 in photogalvanic cell for solar
  energy conversion and storage: Toluidine blue-glucose system},
\newblock \bibinfo{journal}{International Journal of Energy Research}
  \bibinfo{volume}{20} (\bibinfo{year}{1996}) \bibinfo{pages}{581--585}.
  \URLprefix
  \url{http://onlinelibrary.wiley.com/doi/10.1002/(SICI)1099-114X(199607)20:7<581::AID-ER168>3.0.CO;2-4/abstract}.
  \DOIprefix\doi{10.1002/(SICI)1099-114X(199607)20:7<581::AID-ER168>3.0.CO;2-4}.
\bibitem[{Gangotri and Regar(1997)}]{gangotri_use_1997}
\bibinfo{author}{K.~M. Gangotri}, \bibinfo{author}{O.~P. Regar},
\newblock \bibinfo{title}{Use of azine dye as a photosensitizer in solar cells:
  different reductants—safranine systems},
\newblock \bibinfo{journal}{International Journal of Energy Research}
  \bibinfo{volume}{21} (\bibinfo{year}{1997}) \bibinfo{pages}{1345--1350}.
  \URLprefix
  \url{http://onlinelibrary.wiley.com/doi/10.1002/(SICI)1099-114X(199711)21:14<1345::AID-ER356>3.0.CO;2-H/abstract}.
  \DOIprefix\doi{10.1002/(SICI)1099-114X(199711)21:14<1345::AID-ER356>3.0.CO;2-H}.
\bibitem[{Gangotri and Lal(2000)}]{gangotri_studies_2000}
\bibinfo{author}{K.~M. Gangotri}, \bibinfo{author}{C.~Lal},
\newblock \bibinfo{title}{Studies in photogalvanic effect and mixed dyes
  system: {EDTA–methylene} blue + toluidine blue system},
\newblock \bibinfo{journal}{International Journal of Energy Research}
  \bibinfo{volume}{24} (\bibinfo{year}{2000}) \bibinfo{pages}{365--371}.
  \URLprefix
  \url{http://onlinelibrary.wiley.com/doi/10.1002/(SICI)1099-114X(20000325)24:4<365::AID-ER593>3.0.CO;2-I/abstract}.
  \DOIprefix\doi{10.1002/(SICI)1099-114X(20000325)24:4<365::AID-ER593>3.0.CO;2-I}.
\bibitem[{Gangotri and Indora(2010)}]{gangotri_studies_2010}
\bibinfo{author}{K.~Gangotri}, \bibinfo{author}{V.~Indora},
\newblock \bibinfo{title}{Studies in the photogalvanic effect in mixed
  reductants system for solar energy conversion and storage: Dextrose and
  ethylenediaminetetraacetic {acid–Azur} a system},
\newblock \bibinfo{journal}{Solar Energy} \bibinfo{volume}{84}
  (\bibinfo{year}{2010}) \bibinfo{pages}{271--276}. \URLprefix
  \url{http://www.sciencedirect.com/science/article/pii/S0038092X09002722}.
  \DOIprefix\doi{10.1016/j.solener.2009.11.007}.
\bibitem[{Genwa and Genwa(2008)}]{genwa_photogalvanic_2008}
\bibinfo{author}{K.~Genwa}, \bibinfo{author}{M.~Genwa},
\newblock \bibinfo{title}{Photogalvanic cell: A new approach for green and
  sustainable chemistry},
\newblock \bibinfo{journal}{Solar Energy Materials and Solar Cells}
  \bibinfo{volume}{92} (\bibinfo{year}{2008}) \bibinfo{pages}{522--529}.
  \URLprefix
  \url{http://www.sciencedirect.com/science/article/pii/S0927024807004035}.
  \DOIprefix\doi{10.1016/j.solmat.2007.10.010}.
\bibitem[{Genwa and Khatri(2009)}]{genwa_comparative_2009}
\bibinfo{author}{K.~R. Genwa}, \bibinfo{author}{N.~C. Khatri},
\newblock \bibinfo{title}{Comparative study of photosensitizing dyes in
  photogalvanic cells for solar energy conversion and storage:
  Brij-{35−Diethylenetriamine} pentaacetic acid ({DTPA)} system},
\newblock \bibinfo{journal}{Energy \& Fuels} \bibinfo{volume}{23}
  (\bibinfo{year}{2009}) \bibinfo{pages}{1024--1031}. \URLprefix
  \url{http://dx.doi.org/10.1021/ef800747w}. \DOIprefix\doi{10.1021/ef800747w}.
\bibitem[{Lal(2007)}]{lal_use_2007}
\bibinfo{author}{C.~Lal},
\newblock \bibinfo{title}{Use of mixed dyes in a photogalvanic cell for solar
  energy conversion and storage: {EDTA–thionine–azur-B} system},
\newblock \bibinfo{journal}{Journal of Power Sources} \bibinfo{volume}{164}
  (\bibinfo{year}{2007}) \bibinfo{pages}{926--930}. \URLprefix
  \url{http://www.sciencedirect.com/science/article/pii/S0378775306024268}.
  \DOIprefix\doi{10.1016/j.jpowsour.2006.11.020}.
\bibitem[{Bhimwal and Gangotri(2011)}]{bhimwal_comparative_2011}
\bibinfo{author}{M.~K. Bhimwal}, \bibinfo{author}{K.~Gangotri},
\newblock \bibinfo{title}{A comparative study on the performance of
  photogalvanic cells with different photosensitizers for solar energy
  conversion and storage: D-xylose-{NaLS} systems},
\newblock \bibinfo{journal}{Energy} \bibinfo{volume}{36} (\bibinfo{year}{2011})
  \bibinfo{pages}{1324--1331}. \URLprefix
  \url{http://www.sciencedirect.com/science/article/pii/S036054421000633X}.
  \DOIprefix\doi{10.1016/j.energy.2010.11.007}.
\bibitem[{Marcus(1965)}]{marcus_theory_1965}
\bibinfo{author}{R.~A. Marcus},
\newblock \bibinfo{title}{On the theory of {Electron‐Transfer} reactions.
  {VI.} unified treatment for homogeneous and electrode reactions},
\newblock \bibinfo{journal}{The Journal of Chemical Physics}
  \bibinfo{volume}{43} (\bibinfo{year}{1965}) \bibinfo{pages}{679--701}.
  \URLprefix \url{http://jcp.aip.org/resource/1/jcpsa6/v43/i2/p679_s1}.
  \DOIprefix\doi{doi:10.1063/1.1696792}.
\bibitem[{Albery(1982)}]{albery_development_1982}
\bibinfo{author}{W.~J. Albery},
\newblock \bibinfo{title}{Development of photogalvanic cells for solar energy
  conservation},
\newblock \bibinfo{journal}{Accounts of Chemical Research} \bibinfo{volume}{15}
  (\bibinfo{year}{1982}) \bibinfo{pages}{142--148}. \URLprefix
  \url{http://dx.doi.org/10.1021/ar00077a003}.
  \DOIprefix\doi{10.1021/ar00077a003}.
\bibitem[{Bard and Faulkner(2001)}]{bard_electrochemical_2001}
\bibinfo{author}{A.~J. Bard}, \bibinfo{author}{L.~R. Faulkner},
  \bibinfo{title}{Electrochemical methods: fundamentals and applications},
  \bibinfo{publisher}{Wiley}, \bibinfo{address}{New York},
  \bibinfo{year}{2001}.
\bibitem[{John~Albery and Archer(1978)}]{john_albery_photogalvanic_1978}
\bibinfo{author}{W.~John~Albery}, \bibinfo{author}{M.~D. Archer},
\newblock \bibinfo{title}{Photogalvanic cells: Part 4. the maximum power from a
  thin layer cell with differential electrode kinetics},
\newblock \bibinfo{journal}{Journal of Electroanalytical Chemistry and
  Interfacial Electrochemistry} \bibinfo{volume}{86} (\bibinfo{year}{1978})
  \bibinfo{pages}{19--34}. \URLprefix
  \url{http://www.sciencedirect.com/science/article/pii/S0022072878803520}.
  \DOIprefix\doi{10.1016/S0022-0728(78)80352-0}.
\bibitem[{Usgaocar et~al.(2013)Usgaocar, Wang, Mahmoudzadeh, Mirvakili,
  Slota-Newson, Madden, Beatty, and Takshi}]{usgaocar_semiconductors_2013}
\bibinfo{author}{A.~R. Usgaocar}, \bibinfo{author}{L.~Wang},
  \bibinfo{author}{A.~Mahmoudzadeh}, \bibinfo{author}{S.~M. Mirvakili},
  \bibinfo{author}{J.~E. Slota-Newson}, \bibinfo{author}{J.~D. Madden},
  \bibinfo{author}{J.~T. Beatty}, \bibinfo{author}{A.~Takshi},
\newblock \bibinfo{title}{Semiconductors as selective electrodes for
  bio-photovoltaic cells},
\newblock \bibinfo{journal}{Meeting Abstracts} \bibinfo{volume}{{MA2013-01}}
  (\bibinfo{year}{2013}) \bibinfo{pages}{282--282}. \URLprefix
  \url{http://ma.ecsdl.org/content/MA2013-01/4/282}.

\end{thebibliography}
\newpage

\end{document}